\pdfoutput=1
\documentclass[11pt]{article} %for pdfLaTeX
\usepackage{latexsym}
\usepackage{amssymb}
\usepackage{amsmath}
\usepackage{hyperref}
\hypersetup{colorlinks=true, allcolors=blue}
\usepackage{graphicx}% Include figure files

\usepackage{slashed}
\usepackage{mathbbol}

\usepackage[title]{appendix}

\usepackage{color}
\begin{document}
\pagestyle{empty}

\begin{flushright}
KEK-TH-2385
\end{flushright}

\vspace{3cm}

\begin{center}

{\bf\LARGE  
$\mu$TRISTAN
}
\\

\vspace*{1.5cm}
{\large 
Yu Hamada$^{1}$, Ryuichiro Kitano$^{1,3}$, Ryutaro Matsudo$^{1}$,
Hiromasa Takaura$^{1}$\\
and Mitsuhiro Yoshida$^{2,3}$
} \\
\vspace*{0.5cm}

{\it 
$^1$KEK Theory Center, Tsukuba 305-0801,
Japan\\
$^2$KEK Accelerator department, Tsukuba 305-0801, Japan\\
$^3$Graduate University for Advanced Studies (Sokendai), Tsukuba
305-0801, Japan\\

}

\end{center}

\vspace*{1.0cm}

\begin{abstract}
{\normalsize
The ultra-cold muon technology developed for the muon $g-2$ experiment
at J-PARC provides a low emittance $\mu^+$ beam which can be
accelerated and used for realistic collider experiments. We consider
the possibility of new collider experiments by accelerating the
$\mu^+$ beam up to 1~TeV.
Allowing the $\mu^+$ beam to collide with a high intensity $e^-$ beam at the TRISTAN energy, $E_{e^-}= 30$~GeV, 
in the storage ring with the same size as TRISTAN (the circumference of 3 km), one can realize 
a collider experiment with the center-of-mass 
energy $\sqrt s = 346$~GeV,
which allows productions of the Higgs bosons through the vector boson fusion processes.
We estimate the deliverable luminosity with existing accelerator technologies to be at the level of $5 \times 10^{33}$~cm$^{-2}$~s$^{-1}$, with which the collider can be a good Higgs boson factory.
The $\mu^+ \mu^+$ colliders up to $\sqrt s = 2$~TeV are also possible
by using the same storage ring. They have a capability of producing
the superpartner of the muon up to TeV masses.
}
\end{abstract} 

%%%%%%%%%%%%%%%%%%%%%%%%%%%%%%%%%%%%%%%%%%%%%%%%%%%%%%%%%%%%%%%%%%%%%%%%%%%%
\newpage
\baselineskip=18pt
\setcounter{page}{2}
\pagestyle{plain}

\setcounter{footnote}{0}

\section{Introduction}

It is important to perform measurements of the Higgs boson properties
in order to uncover the mysteries of the electroweak symmetry
breaking.
Measurements of the couplings at the level of a percent probe the
energy scale of more than a TeV, where some hints for the origin of our
peculiar vacuum are expected to be hidden.
This motivates us to consider high energy lepton colliders that
copiously produce Higgs bosons.
The $e^+ e^-$ linear colliders such as ILC~\cite{Behnke:2013xla} and
CLIC~\cite{CLICdp:2018cto} can indeed make such measurements possible.
The lowest energy option of the ILC, $\sqrt s = 250$~GeV, will give a
great improvement in the knowledge of fundamental physics.

The $\mu^+ \mu^-$ colliders have also been considered as a future
possibility, where less synchrotron radiation than the case of the
electron enables us to consider a few kilometer-level circular
colliders with TeV energies~(see e.g., Ref.~\cite{Gallardo:1996aa,
Ankenbrandt:1999cta}).
Not only Higgs boson physics, a direct probe of new physics will be
possible with such high-energy colliders. (See
Ref.~\cite{AlAli:2021let} for a recent summary.)

There have been studies of Higgs boson physics at high-energy muon
colliders. Since the cross sections of the vector boson fusion processes
are enhanced at high energies, very precise measurements of Higgs
couplings have been demonstrated to be
possible~\cite{Bartosik:2020xwr, Han:2020pif}.
Those studies, however, assume luminosities which demand further
developments of muon cooling and accelerating
technologies~\cite{Delahaye:2019omf}.

In this paper, we instead consider $\mu^+ e^-$ and $\mu^+ \mu^+$
colliders, which can deliver good enough luminosities for physics
researches within the existing technologies. While a narrow $\mu^-$
beam for muon colliders has not been achieved yet, there is an
established technology to create a low-emittance $\mu^+$ beam by using
the ultra-cold muons~\cite{Kondo:2018rzx}.
At the $\mu^+ e^-$ collider, the Higgs
bosons are produced via the $WW/ZZ$ fusion as in the case of the
$\mu^+ \mu^-$ or $e^+e^-$ colliders while background events through
the $s$-channel annihilation, such as $W^+W^-$ and $q \bar q$ final states, are
absent. Using the storage ring with the circumference of about 3~km
as a reference design, we consider a collider with the
center-of-mass energy, $\sqrt s = 346$~GeV, by accelerating the
electrons and the muons to 30~GeV and 1~TeV, respectively.
Using the same tunnel, it is possible to simultaneously build a $\mu^+ \mu^+$
collider with $\sqrt s = 2$~TeV, which can directly reach TeV physics.

In Refs.~\cite{Lu:2020dkx, Bossi:2020yne}, good physics performances
of $\mu^+ e^-$ colliders have been demonstrated for the measurements
of the Higgs boson couplings as well as searches for lepton flavor
violating interactions by assuming integrated luminosities of the
order of ab$^{-1}$.
It is then an important question whether such a collider can be
constructed in a timely manner.

Two independent projects at KEK have achieved important
accomplishments which actually enable us to plan the construction of
high energy $\mu^+ e^-$ and $\mu^+ \mu^+$ colliders now.
One is the world's highest luminosity at the SuperKEKB experiment, which aims for the instantaneous luminosity of $8\times
10^{35}$~cm$^{-2}$~s$^{-1}$~\cite{Belle-II:2010dht}. 
An upgrade to have a polarized $e^-$ beam is under
consideration~\cite{Roney:2021Bd} that will be important for the Higgs
factory.
Another important technology is the production of the ultra-cold muons, which is
developed for the precise measurements of the $g-2$ of the muon at
J-PARC~\cite{Abe:2019thb}. The positive muons from the pion decay are
stopped at a surface of a material and trap electrons to form muoniums.
By shooting a laser to strip electrons, one can obtain ultra-cold
positive muons, that can be accelerated to be used for a low-emittance
beam for colliders.

Both technologies are already well studied. Considering a similar
proton driver to that in J-PARC and using all of the protons for muon
production, the production rate of the ultra-cold positive muons is
estimated to be at the level of $10^{13-14}$ muons per second. The
ultra-cold technology provides the normalized emittance of the $\mu^+$
beam to be of $4$~mm~mrad~\cite{Abe:2019thb}. By accelerating the $\mu^+$ beam up
to TeV and focusing it at the interaction point, the beam size is reduced
to a few $\mu$m. When we make the $\mu^+$ beam collide with the intense $e^-$ beam at the
TRISTAN energy, 30~GeV, our estimate of the instantaneous luminosity
is of the order of $5 \times 10^{33}$~cm$^{-2}$~s$^{-1}$ per detector,
which is at the same level as the design of the ILC. By assuming ten
years of running with a single detector, the integrated luminosity can
be as high as 1~ab$^{-1}$, with which 0.1 million Higgs bosons are
produced through the $W$ boson fusion process.

A larger storage ring such as the size of the Tevatron allows higher
energy colliders. By assuming that the dipole magnets at about $16$~T
is available by the time of the construction, one can reach $\sqrt s =
775$~GeV with 50~GeV electrons and 3~TeV muons.
The possible luminosity is estimated to be similar to the case of the
1~TeV muons by assuming the same beam power. 
Since the production cross section via the $W$ boson fusion process is
enhanced at high energies, 0.5 million Higgs bosons can be produced
with the same luminosity, 1~ab$^{-1}$. The pair production of
Higgs bosons is also enhanced to be $O(100)$ events, with which one can
expect the measurement of the self-coupling of the Higgs boson.

One can also consider a $\mu^+ \mu^+$ collider~\cite{Heusch:1995yw}
with $\sqrt s = 2$~TeV at the 3~km ring or 6~TeV for the large ring
option.
The instantaneous luminosity is estimated to be of the order of $6
\times 10^{32}$~cm$^{-2}$~s$^{-1}$ by assuming the same $\mu^+$ beam
as above. Although the Higgs production is possible through the $Z$
boson fusion process, the number of the events is much fewer than the $\mu^+
e^-$ colliders due to the limited luminosity and the small $Z$ boson fusion
cross sections.
On the other hand, the $\mu^+ \mu^+$ colliders have a good reach for
new particle searches, such as the pair production of the superpartners
of the muon.

The high intensity $\mu^+$ facility for the $\mu^+ e^-$ or $\mu^+
\mu^+$ colliders will provide rich physics opportunities such as
precision muon physics, muon engineering, material science as well as
neutrino physics. One can also expect future developments towards the
$\mu^+ \mu^-$ colliders or even possibly neutrino colliders. Starting
with a $\mu^+ e^-$ collider may be a good strategic option for future
particle physics.

This paper is organized as follows. In the next section, we discuss
the design of the accelerator complex, and estimate the possible
instantaneous luminosity of the $\mu^+ e^-$ and $\mu^+ \mu^+$
colliders. The Higgs boson productions at the colliders are studied in
Section~\ref{sec:higgs} by using the estimated luminosity, and the
case with the larger ring option is studied in
Section~\ref{sec:largering}. We briefly discuss the requirements for
the detectors in Section~\ref{sec:detector} in order for the Higgs
coupling measurements to be possible. The reaches of the superparticle
searches at $\mu^+ e^-$ and $\mu^+ \mu^+$ colliders are estimated in
Section~\ref{sec:susy}. Section~\ref{sec:summary} is devoted to the
summary.

\section{Ultra-cold muons and $\mu^+$ accelerator}
\label{sec:accelerator}

\begin{figure}[tbp]
  \centering
  \includegraphics[width=\textwidth]{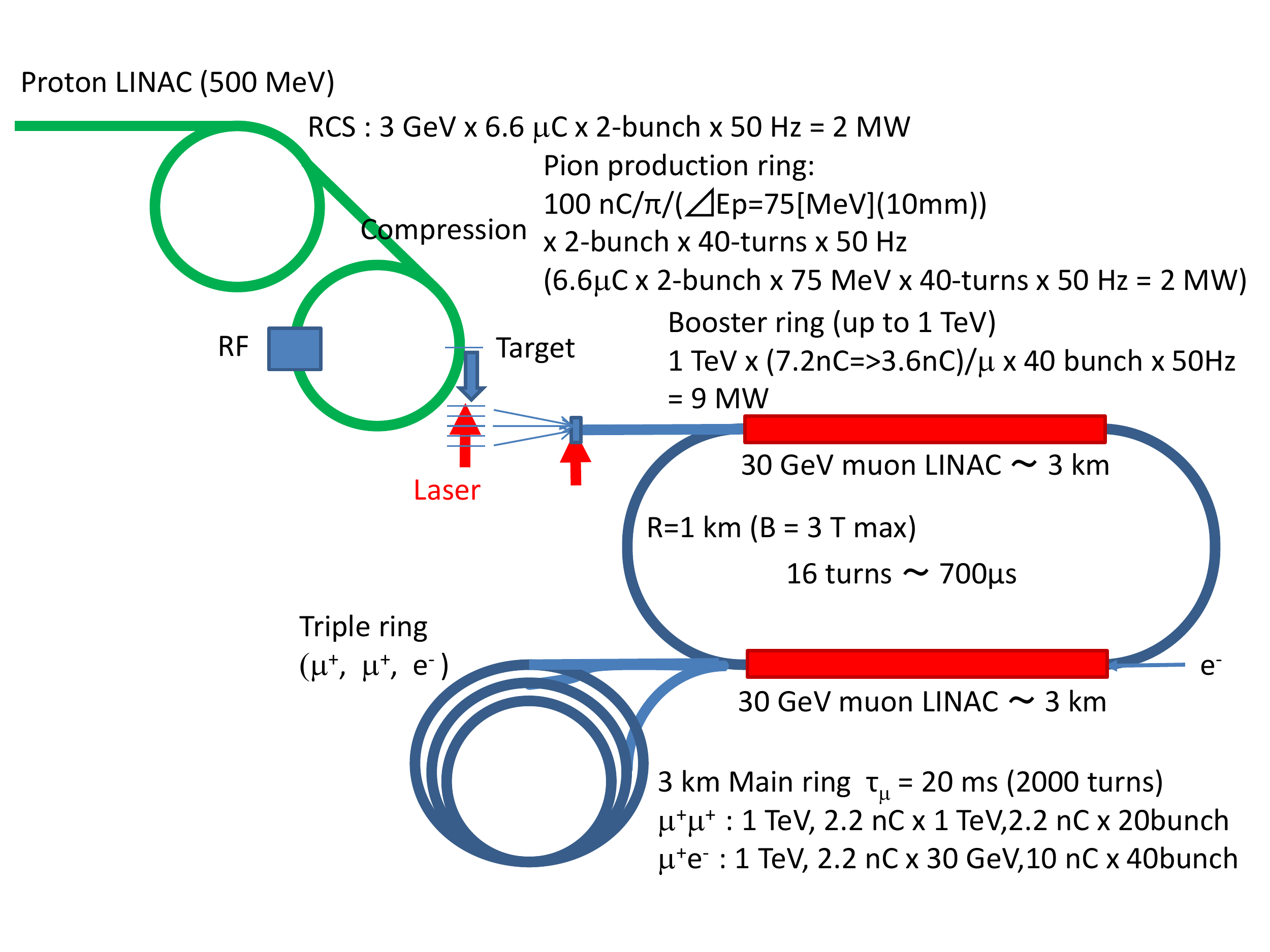}\hspace{1em}
  \vspace{-3em}
  \caption{Conceptual design of the $\mu^+e^-/\mu^+\mu^+$ collider}
\label{fig:acc}
\end{figure}

We discuss the conceptual design of the $\mu^+$ accelerator and the
storage ring. We show the rough sketch in Fig.~\ref{fig:acc}. Based on
this design, we estimate the possible instantaneous luminosity of the
$\mu^+ e^-$ and $\mu^+ \mu^+$ colliders.
The design is based on the muon production and re-acceleration
experiments at the J-PARC Materials and Life Science Experimental
Facility (MLF) but we extended the part of the pion production target
and the pion stopping target so that the proton beam is efficiently
used for the production of ultra-cold muons.
For the technology of the re-acceleration of the ultra-cold muons, see
Ref.~\cite{Kondo:2018rzx}.
The studies of the muon collider~\cite{Gallardo:1996aa} are taken as a
reference for the design concept of the muon acceleration and storage.

%%%%%%%%%%%%%%%%%%%%%%%
%%%%%%%%%%%%%%%%%%%%%%%

%%%%%%%%%%%%%%%%%%%%%%%
\begin{figure}[tbp]
  \centering
  \includegraphics[width=0.8\textwidth]{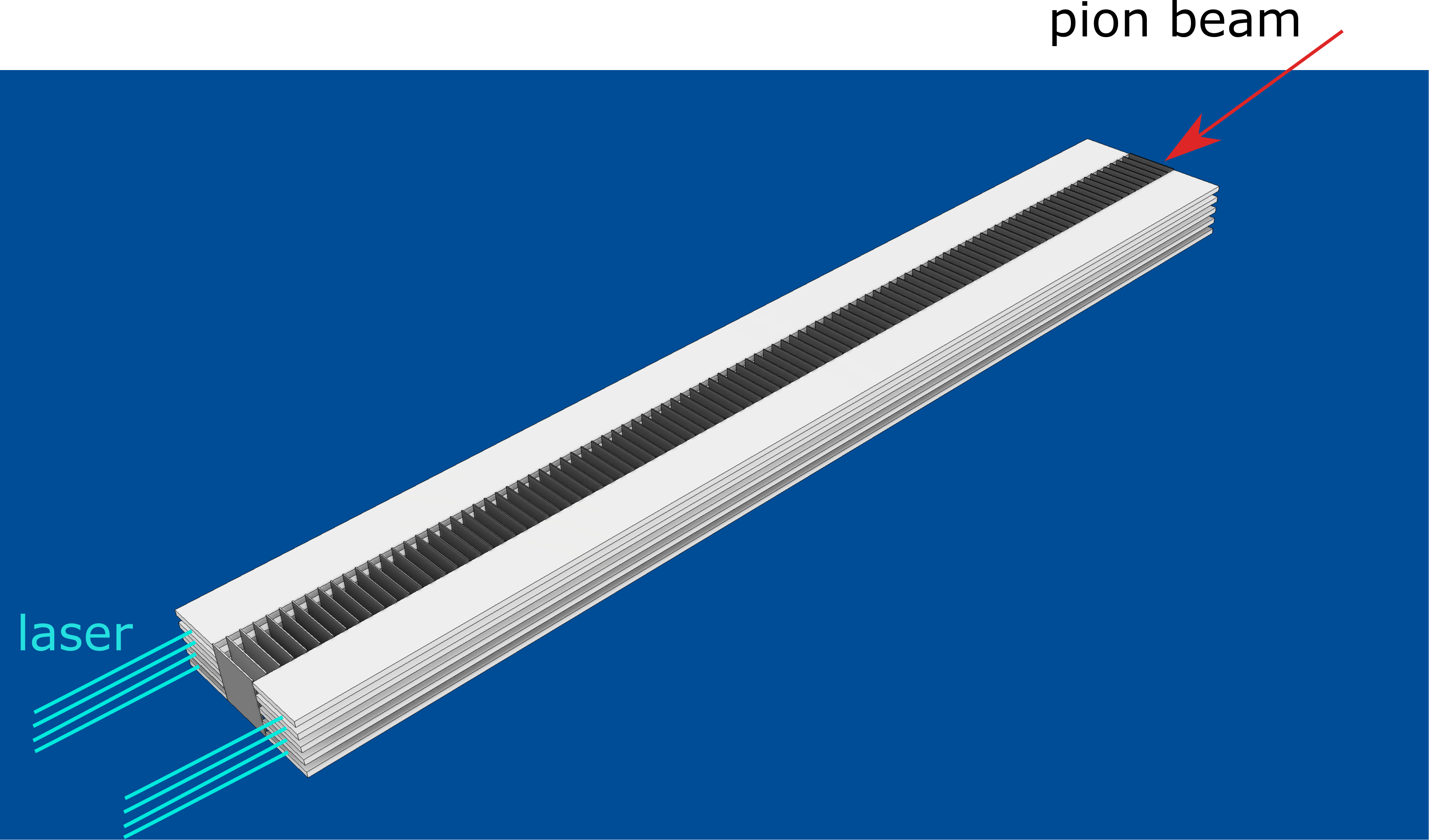}\hspace{1em}
  \caption{The pion stopping target surrounded by silica aerogel for
  muonium production and laser ionization.}
\label{fig:target}
\end{figure}
%%%%%%%%%%%%%%%%%%%%%%%

\subsection{Proton accelerator and pion production ring}

The first stage of the $\mu^+$ production is to deliver an intense
proton beam to the pion production target.
We assume the beam parameters of the proton LINAC (Linear Accelerator)
and RCS (Rapid-Cycling Synchrotron) to be similar to those in J-PARC
with the beam power of 2~MW. The 2-bunch proton beam with the bunch
charge 6.6~$\mu$C ($=4.1 \times 10^{13}$ protons) is accelerated to
3~GeV with the repetition rate of 50~Hz. The operation at 50~Hz is set
by the lifetime of the muon at the energy of 1~TeV, $\gamma \tau_\mu =
20$~ms, so that a new beam is injected after muons spend their
lifetime in the storage ring.

The 3~GeV proton beam is transported to the pion production ring, where
the graphite target with the thickness of 10~mm is placed. The proton
beam is recirculated to repeatedly hit (40 times for each bunch) the
target.
Such a configuration is possible by using the longitudinal phase
rotation with suitable $R_{56}$ and the RF cavity to compensate the
decreased energy by the graphite target inside the ring.
The energy loss of each proton by the collision is estimated to be
75~MeV (2.5~\% energy loss), which amounts to 2~MW power consumption
used for the recovery of the beam energy.

The cross section of the pion production process, $ p + {\rm C}
\rightarrow \pi + X $, is 150~mb at the proton energy of 3 GeV. The
number of generated pions is calculated as 0.016/proton for the target
thickness of 10 mm. It results in the pion charge of $110~$nC/bunch
(times 40 turns times 2 bunches) at the rate of 50~Hz.\footnote{
The proton beam reduces its bunch charge as it hits the graphite target,
since some of the protons turn into pions. 
As a result, the 40 pion bunches do not have the same charge.
In the present study we neglect the reduction of 
the pion charge for simplicity. A detailed study of the proton 
loss and the pion loss is left for a future work. 
We remark, however, that it is possible to realize 
the pion number (or the muon number) given here
simply by raising the initial proton beam power.
}

The produced charged pions mostly travel to the beam direction
and are transported to the large facility of the multi-layer pion
stopping target (Fig.~\ref{fig:target}) which we explain in the next
subsection.
This part is different from the J-PARC configuration, where only the pions
stopping at the production target, which are actually a small part of the produced pions, are used for the muon production.
Instead, we use the majority of the produced pions, which do not stop
at the graphite target. We estimate the efficiency of the
transportation to the stopping target as 50~\% considering the pion
lifetime and the pion loss by interactions with materials.

\subsection{Surface muon production and cooling}

The multi-layered tungsten foil target is used to stop the generated
pion beams. In order to completely stop the pions, we need a thousand
layers of 1~mm foils placed with the interval of
1~cm~\cite{Oset:1992fd}. This requires a large target system such as
10~m length in total. The intervals are necessary to extract surface
muons from the decay of the stopped pions. A conceptual design is
shown in Fig.~\ref{fig:target}.
Positively charged surface muons ($\mu^+$) are transported to the multiple layered 
silica aerogel target (placed just next to the tungsten foils), where
the muons capture electrons to form muoniums (the $\mu^+e^-$ bound
state). The aerogel target needs to have a length of 10~m with a layer
structure in the vertical direction so that one can shoot lasers in
the gaps. The neutral muonium atoms are thermally diffused from the aerogel 
and ionized by the Lyman-$\alpha$ laser. The ionized $\mu^+$ has a very
small momentum spread, which is called the ultra-cold muon.

The efficiency for the production of the ultra-cold muons in total is estimated
to be 14~\% from the following estimated efficiencies: 52~\% for the muonium formation, 60~\% for
the vacuum yield, 60~\% after the loss of muoniums due to the decay, and 73~\% for
the laser ionization\footnote{We thank Cedric Zhang to provide us with
those numbers.}.

The ultra-cold muons spread over 10~m are in turn collected and
transported to the second aerogel target with the size of a few cm.
There, the muonium is formed again and re-ionized by laser.
The two bunches of muons from the pion production ring are combined at the
second muon target by tuning the timing of the laser irradiation. The
efficiency to produce the ultra-cold muons at the second muon target
is assumed as 50~\%. Such  high efficiency should be possible since a
very thin target is used for the uniform muon energy. In total, the
efficiency to obtain the ultra-cold muons from the stopped pions is
6.8~\%.

Combining with the 50~\% efficiency of the pion stopping and the pion
production rate of 0.016/proton in the previous subsection, we obtain
the number of the ultra-cold muons as $5.5\times 10^{-4}$/proton.
This amounts to 7.2~nC ultra-cold muons/bunch by combining two bunches
of $6.6~\mu$C protons. By taking into account 40 turns of the proton
beam in the production ring at the 50~Hz operation, we have $9.0
\times 10^{13}$ muons/s.

Compared with the J-PARC MLF experiments, the number of muons is
enhanced by $O(10^5)$. This is basically due to the improvement in the
efficiency to collect pions and muons. The efficiency at J-PARC MLF
is much less than the ideal situation as it is designed mainly for the
neutron production in the downstream.
In the condition of the J-PARC MLF, the number of stopped pions in the
graphite target of 20~mm length is $1.5 \times 10^{-3}$/proton. 
Among them, those which stopped in the region of the thin surface of
0.5~mm can be used as the surface muons, whose number is $3 \times
10^{-5}$/proton.
The solid angle to cover the muon capture is 108~mSr, that reduces the
number to $3 \times 10^{-7}$/proton, and 1/3 of them can be
transported to the experimental site. In total, the number of muons
which can be used in the experiment is $1 \times 10^{-7}$/proton.
Comparing with our design, $5.5 \times 10^{-4}/$proton, and 40 turns
times 2 bunches of proton beams, one can see the enhancement of
$O(10^5)$.

\subsection{Muon accelerator}

The ultra-cold muons are pre-accelerated and injected to the booster
ring of the race-track shape (see Fig.~\ref{fig:acc}). The length of
the LINAC part is 3~km and there are arc sections with $R=1$~km where
the alternative gradient bending magnet of 3~T is used.
The booster ring is used to accelerate the $\mu^+$ beam up to the
energy of 1~TeV.
In each of the two LINAC parts of the booster ring, the $\mu^+$ beam is
accelerated by 30~GeV, with which 1~TeV beam is achieved by 16 turns.
This takes about 700~$\mu$s.
The LINAC is also used for the electron acceleration
intermittently for the $\mu^+ e^-$ collider.

During the acceleration, about a half of the muons decays and the
bunch charge is reduced to about 3.6~nC.
Taking into account the beam loss, the total beam loading is about
9~MW. 

\subsection{Main ring and the luminosity}

The accelerated beam is transported to the main ring with the
circumference of 3~km. In accordance with the injection rate at 50~Hz,
we keep the muon beam in the main ring for 20~ms, which is equal to
the muon lifetime.
Due to the decay of the muons during this time, the
time-average number of the muons is $(1-1/e) N_{\rm initial}$ with
$N_{\rm initial}=3.6$~nC, which reduces to about 2.3~nC.
In general, for an injection rate at $50/\kappa$~Hz, the reduction
factor of the time-average number is given by $(1 - e^{-\kappa})
/\kappa$. A large $\kappa$ reduces the luminosity linearly while a
small $\kappa$ requires more electric power consumption at the booster
ring. We take $\kappa = 1$ as the optimal choice.

For the $\mu^+\mu^+$ collider, the $\mu^+$ beam is split into two
beams, so that each beam has 20 bunches, while the $\mu^+ e^-$
collider can use all the 40 bunches. The collision frequency, $f_{\rm
rep}$, for the 3~km ring is 100~kHz times the number of bunches, i.e.,

\begin{align}
  f_{\rm rep}^{(\mu^+ e^-)} = 4~{\rm MHz},
  \quad
  f_{\rm rep}^{(\mu^+ \mu^+)} = 2~{\rm MHz},
\end{align}
respectively.

By the ultra-cold muon technology, one can obtain the normalized
emittance, $\beta \gamma \epsilon$, of the $\mu^+$ beam to be 4~mm
mrad~\cite{Abe:2019thb}. By the acceleration up to 1~TeV, i.e., $\beta
\gamma \sim 10^4$, the emittance is reduced to $\epsilon = 420$~nm
mrad.
By taking the realistic values of the beta functions of the collision
point as $\beta_x = 30$~mm and $\beta_y = 7$~mm, we obtain the beam
sizes $\sigma_x$ and $\sigma_y$ as
\begin{align}
  \sigma_x = 3.6~\mu {\rm m}, \quad
  \sigma_y = 1.7~\mu {\rm m}.
\end{align}
We conservatively assume the same focusing of the $e^-$
beam although much better focusing should be possible by comparing
with the design of SuperKEKB.

The number of muons is reduced at the collision point by the decay
of $\mu^+$ in the storage ring as stated above. By
taking 2.3~nC as the bunch charge, the number of $\mu^+$ per bunch as
\begin{align}
  N_{\mu^+} = 1.4 \times 10^{10}.
\end{align}
For the $e^-$ beam, we take 10~nC as a realistic bunch charge, i.e.,
\begin{align}
  N_{e^-} = 6.2 \times 10^{10}.
\end{align}

By using the formula for the instantaneous luminosity,
\begin{align}
  {\cal L} = {N_{\rm beam 1} N_{\rm beam 2} \over 4 \pi \sigma_x \sigma_y} f_{\rm rep},
\end{align}
we obtain the luminosity of the $\mu^+ e^-$ collider as
\begin{align}
  {\cal L}_{\mu^+ e^-} = 4.6 \times 10^{33}~{\rm cm}^{-2}~{\rm s}^{-1}.
  \label{eq:mue_lumi}
\end{align}
In the case of the $\mu^+ \mu^+$ collider, both of the beams reduce
their intensities due to the muon decay. For the operation at 50~Hz, the
time averaged reduction due to the decay is $(1-1/e^2)/2$ rather than
$(1-1/e)^2$.
Noting this, we obtain the luminosity of the $\mu^+ \mu^+$ collider as
\begin{align}
  {\cal L}_{\mu^+ \mu^+} = 5.7 \times 10^{32}~{\rm cm}^{-2}~{\rm s}^{-1}.
  \label{eq:mumu_lumi}
\end{align}
These luminosities in Eqs.~\eqref{eq:mue_lumi} and
\eqref{eq:mumu_lumi} are delivered for each detector if we have
multiple collision points in the storage ring.

Although the precise numbers depend on the various efficiencies for
the muon production as well as the detail designs of the muon
accelerator and the storage ring, the above estimates represent the
realistic orders of magnitudes that one can achieve with currently
available technologies. Detail studies of each component are desired
for more precise estimates.

\subsection{Large ring option}

It is possible to consider the option to have a larger storage ring with a
higher energy. For example, one can consider a 3~TeV muon beam with
the storage ring with the circumference of 9~km. In this case, due to
the longer muon lifetime, the muons can travel larger distances while
a larger ring reduces the collision frequency by 1/3.
The luminosities are, however, kept unchanged since this reduction of
the collision frequency can be compensated by the improved beam
emittance due to the higher energy.
In addition, the acceleration to the higher energy actually requires the same electric power at the booster ring (9~MW) since the repetition rate can be reduced by the same factor due to the longer lifetime.
In total, there is no much gain in terms of luminosities by going to
higher energies unless we increase the beam power.

\subsection{Beam polarizations}

The design of the muon $g-2$ experiment at J-PARC aims at the
polarization of $P_{\mu^+} > 0.9$~\cite{Abe:2019thb}.
The surface muons produced by the decay of $\pi^+$ are 100~\% polarized
due to the $V-A$ structure of the weak interaction. 
Under the magnetic field of order 0.3~T in the longitudinal direction,
the spin of $\mu^+$ is maintained in the formation of the muonium, and
a highly polarized muon beam can be extracted after the laser
ionization.
Although the understanding of the disturbance of the beam emittance by
the magnetic field seems to require more study, we assume in this work
that $P_{\mu^+} = 0.8$ can be obtained. 
At worst, without the longitudinal magnetic field, the formation of
the muonium still leaves the 50~\% polarization at each muonium
production as the $|++\rangle$ muonium states maintains the muon spin
while $|+-\rangle$ states undergo the spin oscillation. The muonium
productions at two targets would reduce the polarization to $P_{\mu^+}
= 0.5^2 = 0.25$.
The difference between $P_{\mu^+} = 0.8$ and 0.25 at $\mu^+ e^-$
colliders results in 30~\% decrease in the cross sections of the $WW$
fusion process for the Higgs boson production. In the case of the
$\mu^+ \mu^+$ collider, the cross sections of purely left-handed muon
process, such as the productions of supersymmetric particles discussed
below, will be reduced by a factor of two.

The beam polarization option of the $e^-$ beam at the SuperKEKB
experiment has been studied~\cite{Roney:2021Bd}.
The polarized electron source and spin rotators placed right before
and after the interaction region(s) have been considered, with the
targeted polarization of $P_{e-} = \pm 0.7$. We assume that the same
technology can be implemented in the $\mu^+ e^-$ colliders.

\subsection{Magnet design}

As the reference design, we assumed the circumference of the storage
ring as the size of the TRISTAN ring, 3~km. The electron energy of
30~GeV is the maximum energy reached in the TRISTAN experiments. For
steering of the 1~TeV muon beam within the 3~km ring, we need a dipole
magnet with the magnetic field of $10$~T, which is a similar strength
as the design of the high-luminosity LHC experiments of 11~T. The
prototypes of utilizing the Nb$_3$Sn superconducting technologies have
already been produced and tested~\cite{Bordini:2019fan}.

If we assume that a larger magnetic field such as 16~T is available by
then, a smaller storage ring is possible. For example, the large ring
option of the 9~km circumference can be reduced to about 6~km although
we still need a large booster ring for the acceleration. The reduced
one can certainly fit realistic experimental sites such as at the
Tevatron site ($\mu$Tevatron).

\section{Higgs boson production at $\mu$TRISTAN}
\label{sec:higgs}

Based on the estimated luminosity in the previous section, we discuss
the rate of the Higgs boson production at the $\mu^+ e^-$ collider
with the energies $(E_{e^-}, E_{\mu^+}) = (30~{\rm GeV}, 1~{\rm TeV})$.
We assume that the beam polarizations with $P_{e^-} = \pm
0.7$ and $P_{\mu^+} = \pm 0.8$ are possible. 
We set the polarizations to be  $(P_{e^-}, P_{\mu^+})=(-0.7,0.8)$,
which maximizes the cross section of the $W$ boson fusion process.

Using these parameters, the cross sections of the single Higgs boson
production via the $W$ boson fusion (WBF) and the $Z$ boson fusion
(ZBF) processes, respectively, are given by 
\begin{align}
\sigma_{\rm WBF} \approx 91~{\rm fb},\quad{} \sigma_{\rm ZBF} \approx 4~{\rm fb} .
\label{eq:sigma_emu}
\end{align}
These cross sections and the energy and angular distributions of the
Higgs boson productions discussed below are all those at the parton
level, and they are calculated by using the
MadGraph5~\cite{Alwall:2014hca}.

The integrated luminosity of 1~{ab}$^{-1}$ can be achieved by ten
years of running by assuming the duty factor of 70~\% and a single
detector.
With this integrated luminosity, the number of Higgs boson events is
then estimated to be
\begin{align}
N( {\rm Higgs})
%&=9.4 \times 10^{4}~[{\rm ab}] \times 4.5 [{\rm ab}^{-1}] \times \left( \frac{{\text{integrated luminosity}}}{4.5 [{\rm ab}^{-1}]} \right) \\ \nonumber
&=9.5 \times 10^4  \times \left( \frac{{\text{integrated luminosity}}}{1.0~{\rm ab}^{-1}} \right) .
\end{align}

This huge number of Higgs events can be used for the precise
measurements of the Higgs boson couplings.
In the $\kappa$ scheme, where $\kappa$ parameters are multiplied to
the Standard Model couplings, the number of the Higgs production via
WBF followed by the $H \to b \bar b$ decay is proportional to
$\kappa_W^2 \kappa_b^2/\kappa_H^2$, where $\kappa_W$ and $\kappa_b$ are the Higgs
boson coupling to the $W$ boson and the $b$ quark normalized by the
Standard Model values. $\kappa_H$ is defined as
$\Gamma_H=\kappa_H^2 \Gamma_H^{\rm SM}$, where 
$\Gamma_H$ denotes the total decay width of the Higgs boson.
Comparing this prediction with the experimental
data, one can restrict the deviations from unity.
The statistical uncertainties for the deviation is estimated to be 
\begin{align}
\Delta (\kappa_W+\kappa_b-\kappa_H)_{\rm stat}
&=\frac{1}{2} \frac{1}{\sqrt{N({\rm WBF}) \times {\rm Br}(h \to b\bar{b}) \times {\rm efficiency}}}
 \nonumber\\
&=3.1 \times 10^{-3} \times \left( \frac{{\text{integrated luminosity}}}{1.0~{\rm ab}^{-1}} \right)^{-1/2}
 \left(\frac{{\text{efficiency}}}{0.5} \right)^{-1/2}.
\label{eq:delta_kappa}
\end{align}
Here the branching ratio, ${\rm Br}(h \to b \bar{b})=5.82 \times
10^{-1}$, is used, and it is assumed that the number of the background
events gets much less than that of the signal events by appropriate
selection cuts. The efficiency is assumed to be 50\%, which actually
depends on the energy resolutions and event selections.
As a reference, in Ref.~\cite{Durig:2014lfa} the event selections at
the ILC at $\sqrt s = 250~{\rm GeV}$ and 500~GeV are considered, which
gives the efficiencies of order 15\% and 40\%, respectively.
In the $\mu^+ e^-$ colliders, there are no processes with the final
state $W^+ W^-$, $ZZ$, $q \bar q$ or $Zh$, which significantly
contribute to background events, and thus the event selection would be
able to be much looser.

The main background is the $Z$ boson production followed by the decay into bottom quark
pair,
$e^{-} \mu^+ \to \nu_e \bar{\nu}_{\mu} Z  \to \nu_e \bar{\nu}_{\mu} b
\bar{b}$.
The invariant mass of the two $b$ jets is peaked at the $Z$ boson
mass, but give an overlap in the Higgs mass region by the finite
energy resolutions of the detector. The number of the events is at
the same level as those of the signals:  
\begin{equation}
N_{\rm BG}= 5.5 \times 10^4 \times \left( \frac{{\text{integrated luminosity}}}{1.0~{\rm ab}^{-1}} \right) .
\end{equation}
The selection cut on the $b \bar b$ invariant mass should
significantly reduce this background. For example, the study in
Ref.~\cite{Durig:2014lfa} have reported that 95\% (90\%) of the $Z \to
b \bar b$ background events are rejected by the cut at $\sqrt s =
250~{\rm GeV}$ $(500~{\rm GeV})$ while about 80\% of the signal events
survive.   
We therefore anticipate that the efficiency of 50\%, assumed e.g. in
Eq.~\eqref{eq:delta_kappa}, provides reasonable estimates.

The study in Ref.~\cite{Lu:2020dkx} reported the precision of the
Higgs coupling to be a percent level by performing a fast detector
simulation.
A better sensitivity in Eq.~\eqref{eq:delta_kappa} is partly due to
enhanced cross sections by the beam polarization. One should,
however, study the detector performance to obtain a realistic
estimate. As we discuss in Section~\ref{sec:detector}, the coverage
and the performance in the forward region (the $\mu^+$ direction) is
especially important.

Decay modes other than $h \to b \bar b$ and the Higgs production via ZBF should also be able to be used
for the measurements of the coupling constants and would be useful in
disentangling individual couplings such as $\kappa_W$ and $\kappa_b$
rather than its combinations.  
Compared to the $e^+ e^-$ or $\mu^+ \mu^-$ colliders, the domination of the WBF process
for the Higgs production is a limitation for this respect, although the
measurements of the subdominant ZBF processes may give an interesting input. See
Refs.~\cite{Abramowicz:2016zbo,Bambade:2019fyw,deBlas:2019rxi} 
for the studies of the Higgs coupling measurements at future
colliders. 
It is important to perform a thorough study of the performance of this
collider for the Higgs coupling measurements to compare with other
future colliders. We will leave such studies for future works as we
need to fix the detector design to proceed. Our first estimate of the
number of the Higgs events is indicating that the $\mu$TRISTAN can
potentially be an interesting realistic option for the near future.

The $\mu^+ \mu^+$ collider at $\sqrt s = 2$~TeV can also produce the
Higgs boson via the $Z$ boson fusion process. The cross section of the
Higgs production at $\sqrt s = 2$~TeV is estimated to be about 54~fb
by assuming the polarized beam $P_{\mu^+} = 0.8$ for both of $\mu^+$.
Although the vector boson fusion process is 
enhanced by a logarithmic factor, $\log s$, 
the small coupling between the lepton and the $Z$ boson results in
a smaller cross section than $\sigma_{\rm WBF} + \sigma_{\rm ZBF}$ in
Eq.~\eqref{eq:sigma_emu}.
Since the luminosity is expected to be reduced by about a factor of
ten compared to $\mu^+ e^-$ colliders, the precision measurements of
the Higgs couplings are better performed at the low energy $\mu^+ e^-$
option.
Even with the reduced luminosity, the $\mu^+ \mu^+$ collider has a
much better capability of supersymmetry searches as we discuss later.

\section{$\mu$Tevatron option}
\label{sec:largering}

The option with a larger storage ring, such as the size of the
Tevatron (the circumference of about 6~km), can provide higher energy
beams. By assuming the same luminosities, one can achieve better
precision of the Higgs boson couplings. The pair production of
Higgs bosons has a large cross section, which makes it possible to
measure the three-point self coupling of the Higgs boson. We assume
that the electron energy of 50~GeV and the muon beam energy of 3~TeV
are possible by using the improved bending magnet (or having 9~km ring).

The cross section of the single Higgs boson production is enhanced as
the energy increases. At $(E_{e^-}, E_{\mu^+}) = (50~{\rm GeV}, 3~{\rm
TeV})$, which gives $\sqrt s = 775~{\rm GeV}$, the WBF and ZBF cross
sections are given by
\begin{align}
\sigma_{\rm WBF} \approx 472~{\rm fb},\quad{} \sigma_{\rm ZBF} \approx 20~{\rm fb} .
\end{align} 
The number of the Higgs boson production events is estimated as
\begin{align}
N( {\rm higgs})
%&=4.9 \times 10^{5}~[{\rm ab}] \times 4.5 [{\rm ab}^{-1}] \times \left( \frac{{\text{integrated luminosity}}}{4.5 [{\rm ab}^{-1}]} \right) \\ \nonumber
&=4.9 \times 10^5  \times \left( \frac{{\text{integrated luminosity}}}{1.0~{\rm ab}^{-1}} \right) ,
\end{align}
and the precision of the $\kappa$ parameters is
\begin{align}
\Delta (\kappa_W+\kappa_b-\kappa_H)_{\rm stat}
%&=\frac{1}{2} \frac{1}{\sqrt{N({\rm WBF}) \times {\rm Br}(h \to b\bar{b}) \times {\rm efficiency}}} \\ \nonumber
&=1.3 \times 10^{-3} \times 
\left( \frac{{\text{integrated luminosity}}}
{1.0~{\rm ab}^{-1}} \right)^{-1/2} 
\left(\frac{{\text{efficiency}}}{0.5} \right)^{-1/2} .
\label{eq:kappa_highE}
\end{align}

The cross section of the Higgs boson pair production process $e^-
\mu^+ \to \nu_e \bar{\nu}_{\mu} h h$ is given by
\begin{equation}
\sigma \approx 8.9 \times 10^{-2} ~{\rm fb} .
\end{equation}
We find the number of events to be
\begin{align}
N( {\text{Higgs pair}})
&=89  \times \left( \frac{{\text{integrated luminosity}}}{1.0~{\rm ab}^{-1}} \right) .
\end{align}
With this number, one should be able to measure the three-point
coupling at the level of ten to a hundred percent.

The good precision of the single Higgs production in
Eq.~\eqref{eq:kappa_highE} is already sensitive to the radiative
corrections to the Higgs coupling~\cite{McCullough:2013rea}. The
effects of the three-point Higgs coupling to the WBF process at the
one-loop level is found to be $0.006
\kappa_\lambda$~\cite{DiVita:2017vrr}, where $\kappa_\lambda$ is the
three point coupling normalized by the Standard Model prediction.
This means that the measurement of $\Delta \kappa_\lambda$ of $\mathcal{O}(20\%)$ is already possible with the single Higgs production measurements.

\section{Asymmetric $\mu^+ e^-$ colliders and detector design}
\label{sec:detector}

The asymmetry in the energies of the colliding beams, 30~GeV and 1~TeV,
or 50~GeV and 3~TeV, makes the final state particles to be boosted to
the direction of the $\mu^+$ beam. The detector should be designed
such that the small-angle region from the beam direction is covered.

\begin{figure}[tbp]
  \centering
  \includegraphics[width=0.47\textwidth]{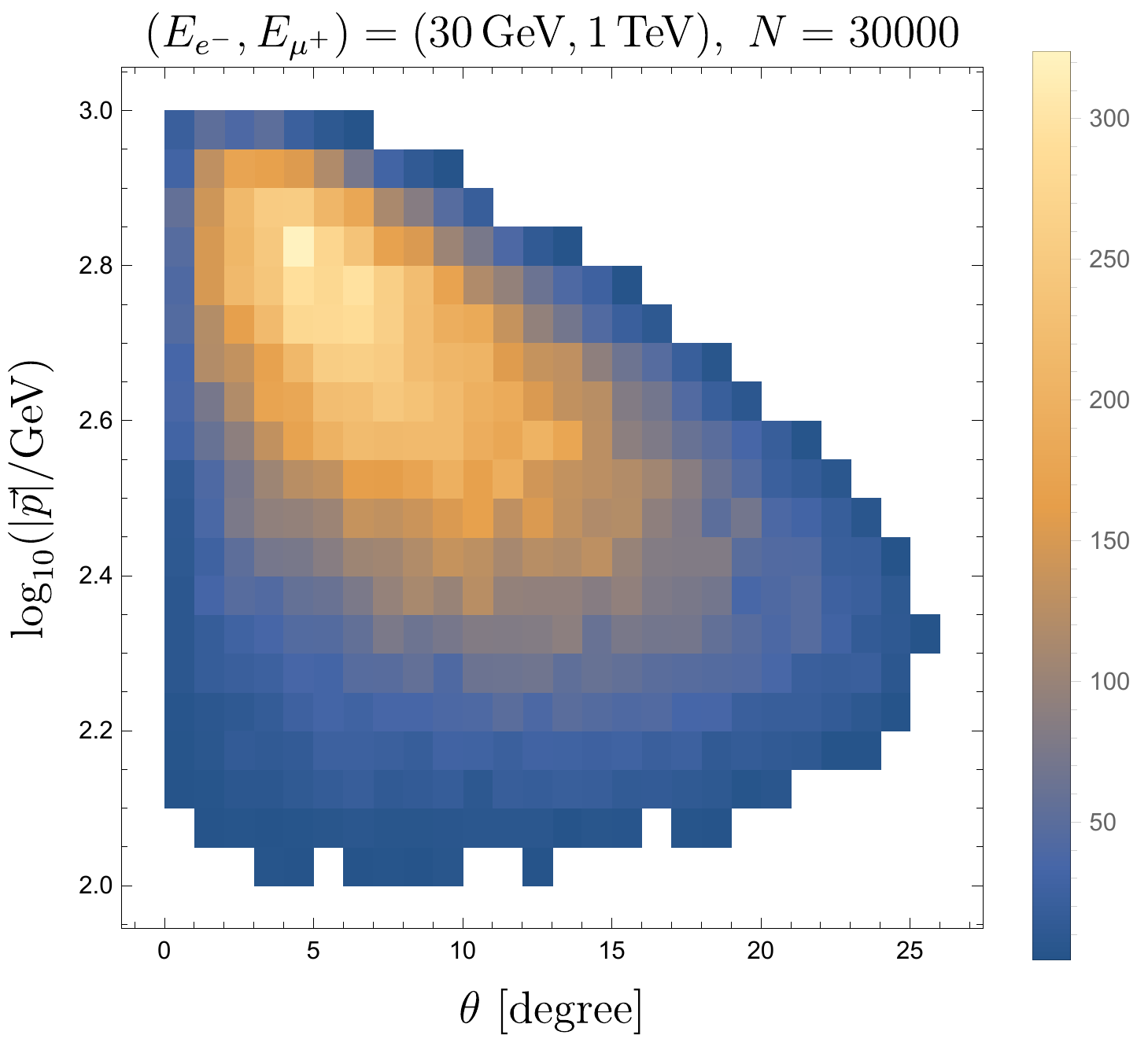}\hspace{1em}
  \includegraphics[width=0.47\textwidth]{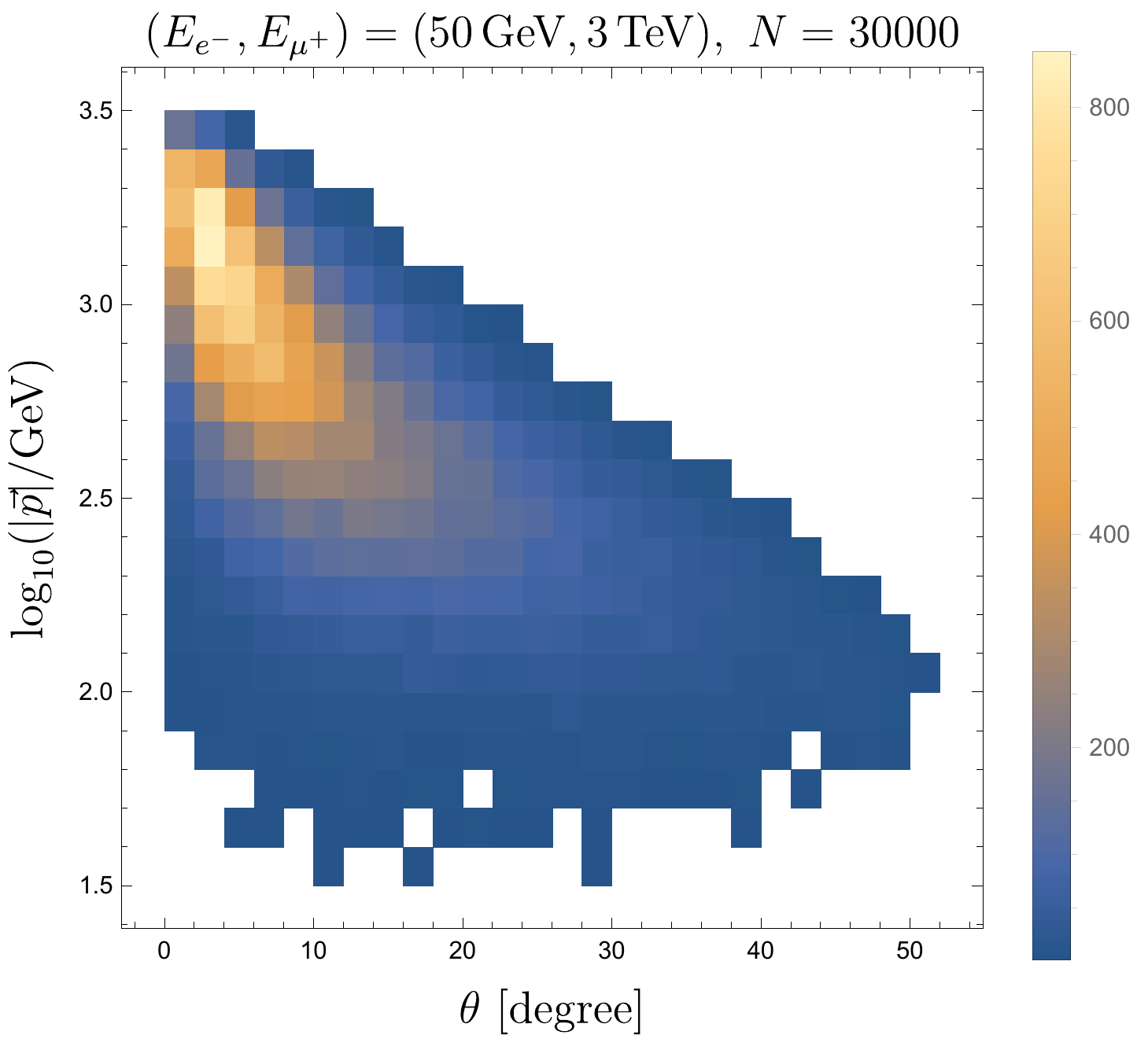}
  \caption{Density histogram plots of Higgs boson produced by the
    $W$-fusion. The number of the event is $N=30000$. The vertical and
    horizontal axes represent the 3D momentum and angle from the beam
    axis, respectively. (Left): The beam energy is $(E_{e^-},
    E_{\mu^+})=(30~{\rm GeV}, 1~{\rm TeV})$. (Right): The beam energy
    is $(E_{e^-}, E_{\mu^+})=(50~{\rm GeV}, 3~{\rm TeV})$.}
\label{fig:density_h}
\end{figure}

\begin{figure}[tbp]
  \centering
  \includegraphics[width=0.47\textwidth]{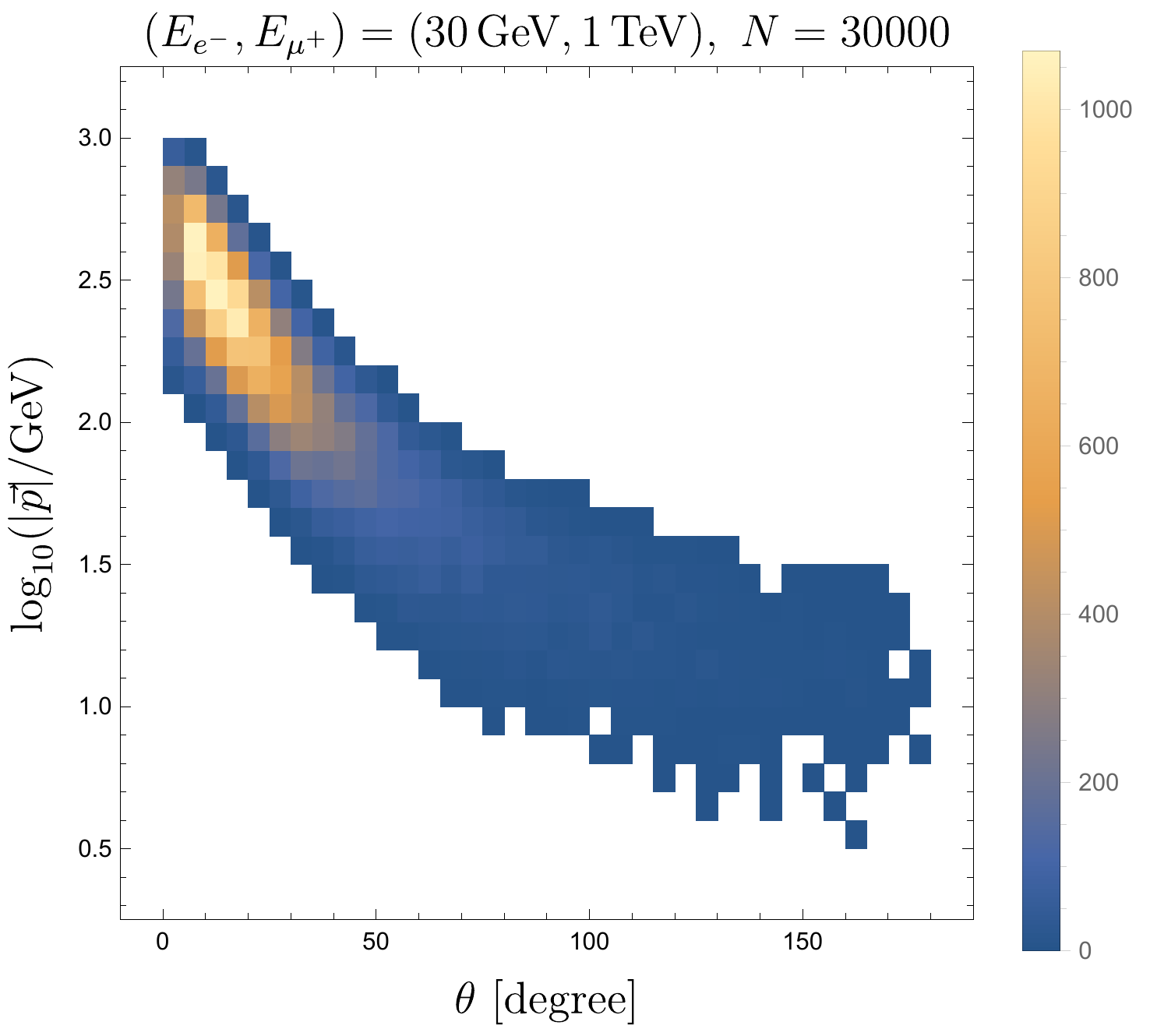}\hspace{1em}
  \includegraphics[width=0.47\textwidth]{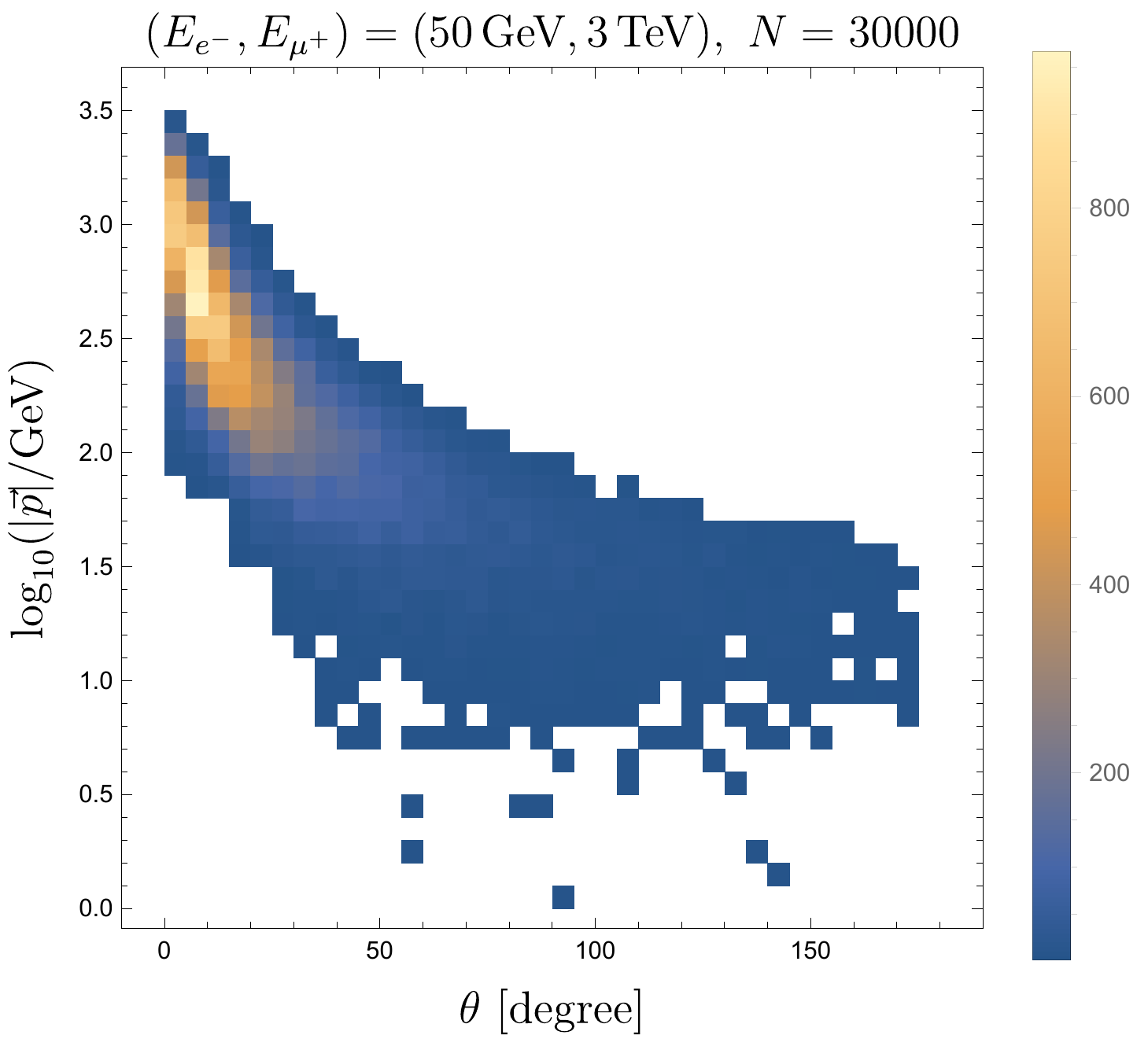}
  \caption{Density histogram plots of $b$ quark produced by the decay
    channel $ h \to b \bar b$. The number of the event is $N=30000$.
    The vertical and horizontal axes represent the 3D momentum and
    angle from the beam axis, respectively. (Left): The beam energy is
    $(E_{e^-}, E_{\mu^+})=(30~{\rm GeV}, 1~{\rm TeV})$. The
    distribution peaks around $|\vec{p}| \simeq 316 \,\mathrm{GeV}$
    and $\theta \simeq 10^\circ$. (Right): The beam energy is
    $(E_{e^-}, E_{\mu^+})=(50~{\rm GeV}, 3~{\rm TeV})$. The
    distribution peaks around $|\vec{p}| \simeq 500 \,\mathrm{GeV}$
    and $\theta \simeq 7.5^\circ$.}
  \label{fig:density_b}
\end{figure}

\begin{figure}[tbp]
  \centering
  \includegraphics[width=0.47\textwidth]{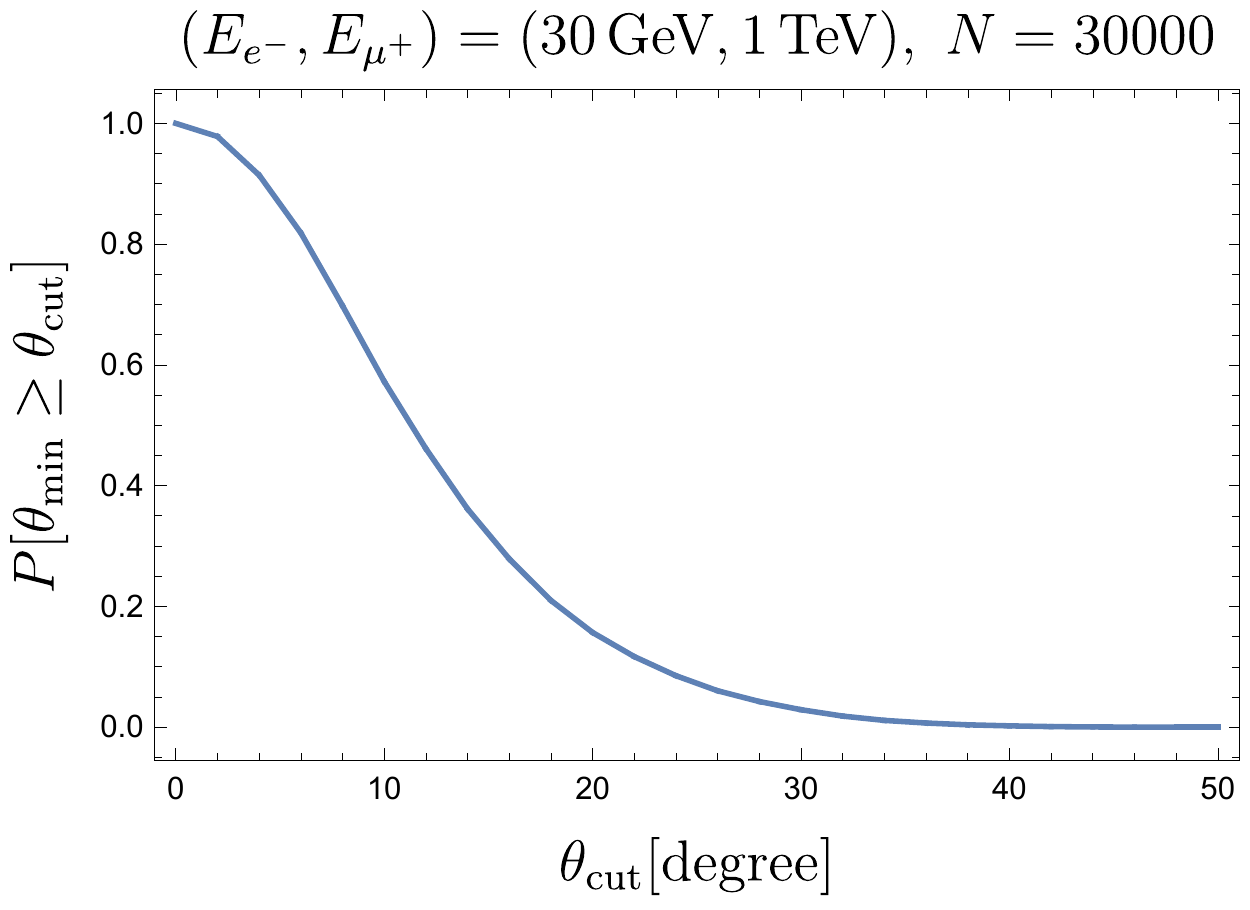}\hspace{1em}
  \includegraphics[width=0.47\textwidth]{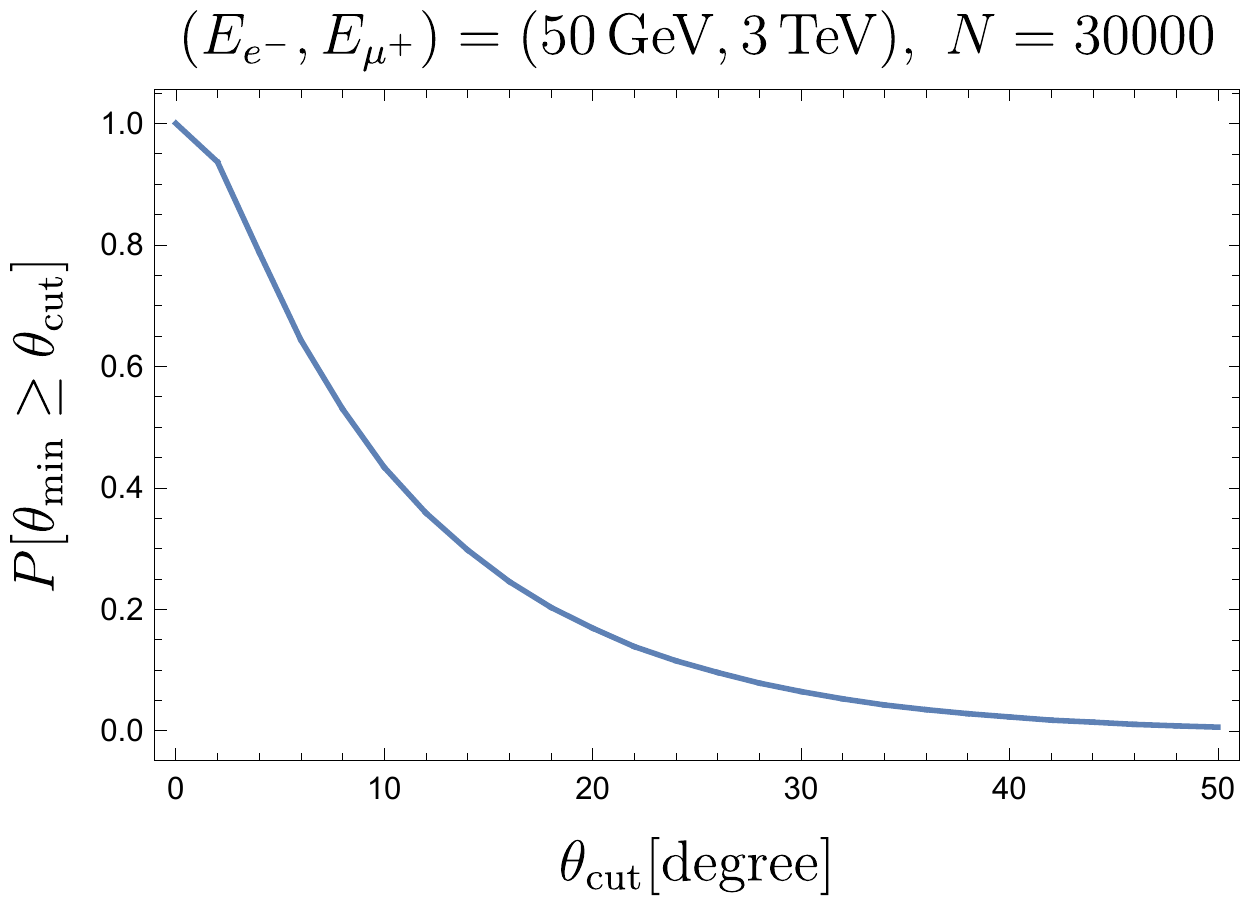}
  \caption{Plots for the number distribution of events such that $\theta_\mathrm{min} \equiv \mathrm{min}\{\theta_b, \theta_{\bar b}\}$ satisfies
    $\theta_\mathrm{min} \geq
    \theta_\mathrm{cut}$. They are normalized by the number of the
    entire events, $N=30000$. (Left): The beam energy is $(E_{e^-},
    E_{\mu^+})=(30~{\rm GeV},1~{\rm TeV})$.
    (Right): The beam energy is $(E_{e^-},
    E_{\mu^+})=(50~{\rm GeV}, 3~{\rm TeV})$.
    In both cases, one needs $\theta_{\rm cut}$ to be less than a few degree for detecting 90~\% of $h \to b \bar b$ events.}
  \label{fig:theta_cut}
\end{figure}

The boost factors from the center-of-mass frame are $\gamma = 3.0$ and
3.9, respectively, for the Higgs boson produced at colliders with energies (30~GeV, 1~TeV) and
(50~GeV, 3~TeV).
This makes the typical polar angle of the final state particles to be
a few to ten degrees in the lab frame.
In Fig.~\ref{fig:density_h}, we show distribution of the polar angle
and the momentum of the Higgs boson produced by the $WW$ fusion
process. We see that the distribution is peaked at $\theta \sim 5-10$
degrees. The $b$ and $\bar b$ quarks from the decay of the
Higgs bosons have the distributions shown in Fig.~\ref{fig:density_b}. 
The fraction of the events where both $b$ and $\bar b$ are in the direction of $\theta > \theta_{\rm cut}$, is shown in Fig.~\ref{fig:theta_cut}.
In order to accept significant fractions of the events, we need a good
detector coverage in the forward direction, such as below a few
degrees.
This requirement may be a challenge in designing detectors, but it
should not be too severe by comparing with the coverage of the forward
region in the LHCb detector ($\theta > 15$~mrad)~\cite{LHCb:2008vvz} or
those of the designs of the electron-proton~\cite{LHeC:2020van} and
electron-ion colliders~\cite{Accardi:2012qut}.

%%%%%%%%%%%%%%%%%%%%
%%%%%%%%%%%%%%%%%%%%
  
%%%%%%%%%%%%%%%%%%%%
%%%%%%%%%%%%%%%%%%%%

%%%%%%%%%%%%%%%%%%%%
%%%%%%%%%%%%%%%%%%%%

In the design of detectors, one should also take into account the
background from the decay products of the beam
muons~\cite{Bartosik:2020xwr}. One should shield the interaction
region from the positrons and photons in the $\mu^+$ beam, and should
also have good efficiency to reject unwanted tracks. Although
placing a shield from the $\mu^+$ beam reduces the $\eta$ coverage in
the direction of the $e^-$ beam, physics performance would not be
significantly affected since the asymmetry in the beam energies makes
the most of the physics events boosted in the $\mu^+$ direction as we
discussed above.
In any case, dedicated studies on the detector designs are necessary
to discuss the physics performance of the $\mu^+ e^-$ colliders.

\section{Superparticle searches at $\mu^+ e^-$ and $\mu^+ \mu^+$ colliders}
\label{sec:susy}

The $\mu^+ e^-$ and also $\mu^+ \mu^+$ colliders have a capability of
producing new particles if kinematically accessible. As an example, we
demonstrate the case with supersymmetry where scalar leptons are
produced through the diagrams of the exchange of $SU(2)_{\rm L}$
gauginos, the Winos. For simplicity, we ignore the Bino (the partner
of $U(1)_{\rm Y}$ gauge boson) and the Higgsino contributions, and we
do not consider decays of the scalar leptons, and just count the
number of production events with luminosities estimated in
Section~\ref{sec:accelerator}. 
In the case where the lightest superparticle is one of the
neutralinos, the final states are $e^- \mu^+$ and missing energy. The
threshold scan of the collider energy (and/or the polarization
dependence of the event rate) should be able to detect the signal if
the number of the events is large enough. We will demonstrate the
reaches by requiring $O(100)$ events per year.
In the actual model, for example, in the case of the Bino as the
lightest supersymmetric particle, one should also include the
production process through the exchange of the Bino. Since the
discussion will be complicated when we include all the supersymmetric
particles, we ignore those amplitudes for simplicity.

The LHC experiments have searched for the events of the pair
production of the scalar leptons, and have excluded the region of the
scalar lepton masses less than about 700~GeV~\cite{ATLAS:2019lff,
CMS:2020bfa} when the mass difference between the scalar lepton and
the lightest neutralino is larger than about 100~GeV.
The center of mass energy of the $\mu^+ e^-$ collider, $\sqrt s =
346$~GeV or 775~GeV, would not allow to go beyond the LHC limits in
terms of the scalar lepton masses. It may, however, cover the region
of the smaller mass differences. The $\mu^+ \mu^+$ colliders, on the
other hand, can reach TeV scale scalar leptons directly as we discuss
below.

\subsection{$e^- + \mu^+ \to \tilde e^- + \tilde \mu^+$}

%%%%%%%%%%%%%%%%%%%%
\begin{figure}[tbp]
  \centering
  \includegraphics[width=0.47\textwidth]{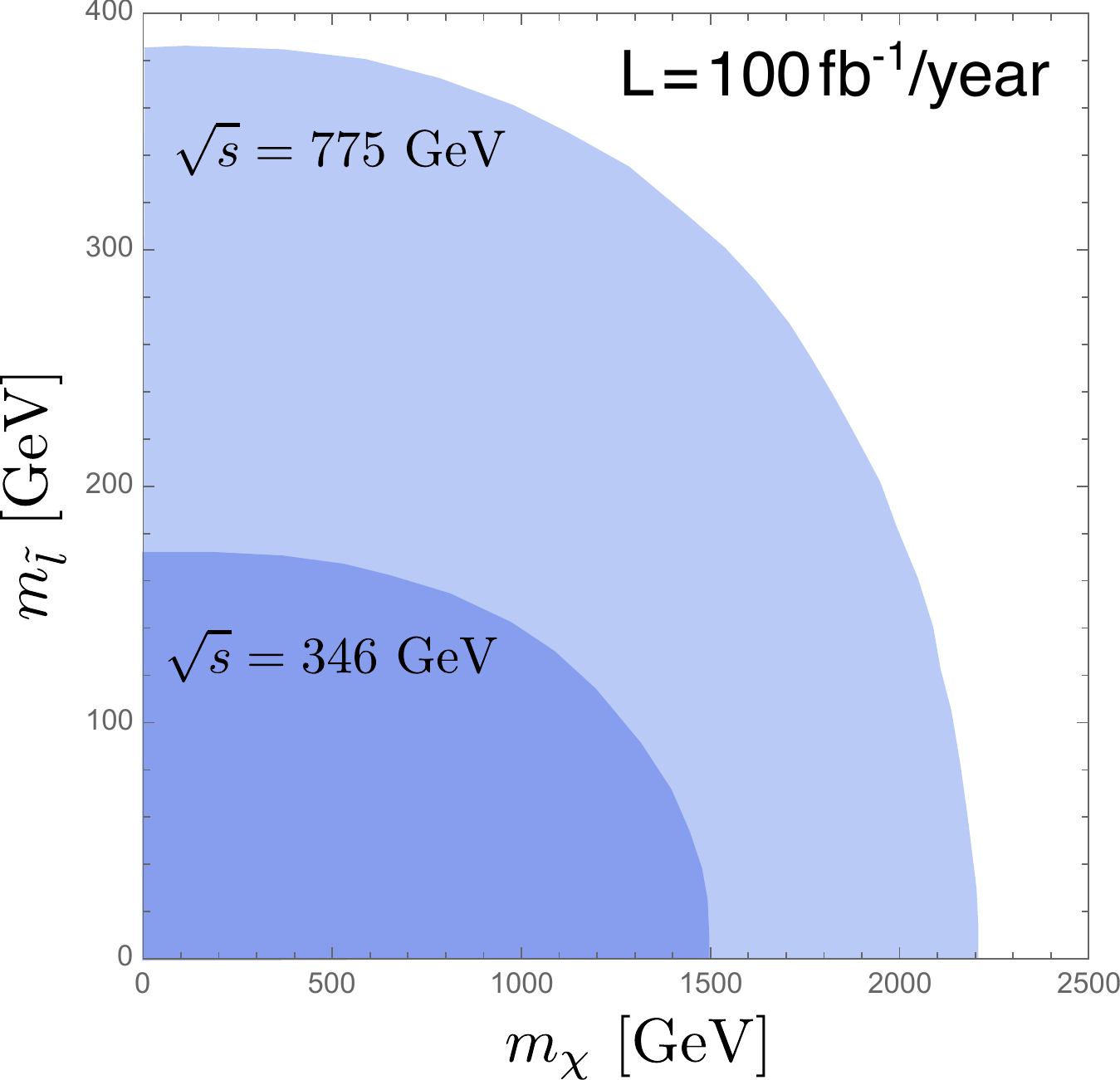}
\caption{The region where the number of the charged slepton pair
production per year is larger than a hundred at the $\mu^+ e^-$
collider. We take the luminosity to be ${\cal L} = 100~{\rm
fb}^{-1}/{\rm year}$. The beam polarizations are taken to be $P_{e^-}
= -0.7$ and $P_{\mu^+} = 0.8$. The scalar lepton masses are taken to
be the same, $m_{\tilde l} = m_{\tilde e} = m_{\tilde \mu}$.}
\label{fig:susy_emu}
\end{figure}
%%%%%%%%%%%%%%%%%%%%

At the $\mu^+ e^-$ colliders, the diagram with neutralino exchange
gives the production of a scalar electron and a scalar muon. 
The differential cross section can be obtained as
\begin{align}
  d \sigma = {d \cos \theta \over 32 \pi} {(1+x_3 - x_4)\beta \over s} | M_{\rm LR} |^2 
  {(1 - P_{e^-}) (1 + P_{\mu^+}) \over 4},
   \quad -1 \leq \cos \theta \leq 1,
\end{align}
where
\begin{align}
  M_{\rm LR} = - {g_2^2 \over 2} \cdot {(1 + x_3 - x_4)\beta \sin \theta \over
  1 + 2 x_A - x_3 - x_4  - (1 + x_3 - x_4) \beta \cos \theta
  },
\end{align}
and
\begin{align}
  x_A = {m_\chi^2 \over s}, \quad
  x_3 = {m_{\tilde e}^2 \over s}, \quad
  x_4 = {m_{\tilde \mu}^2 \over s}, \quad
  \beta = {\sqrt {
1 - 2 x_3 - 2 x_4  + (x_3 - x_4)^2
  }
  \over 1 + x_3 - x_4}.
\end{align}
The masses, $m_\chi$, $m_{\tilde e}$ and $m_{\tilde \mu}$ are those of
the neutralino, the scalar electron and the scalar muon, respectively.
The coupling constant $g_2$ is that of the ${SU(2)_{\rm L}}$ gauge
interaction. A similar (a factor of four larger) cross section is
obtained for scalar neutrino productions.

By setting the maximal optimization of the polarizations, $P_{e^-} =
-0.7$ and $P_{\mu^+} = 0.8$ and assuming the yearly luminosity of
100~fb$^{-1}$ (corresponding to the 70~\% duty factor with
Eq.~\eqref{eq:mue_lumi}), we show in Fig.~\ref{fig:susy_emu} the
parameter region where the number of the production events is over one
hundred per year. In the figure, we take the scalar electron and the
scalar muon to be the same mass, $m_{\tilde l} = m_{\tilde e} =
m_{\tilde \mu}$.

\subsection{$\mu^+ + \mu^+ \to \tilde \mu^+ + \tilde \mu^+$}

%%%%%%%%%%%%%%%%%%%%
\begin{figure}[tbp]
  \centering
  \includegraphics[width=0.47\textwidth]{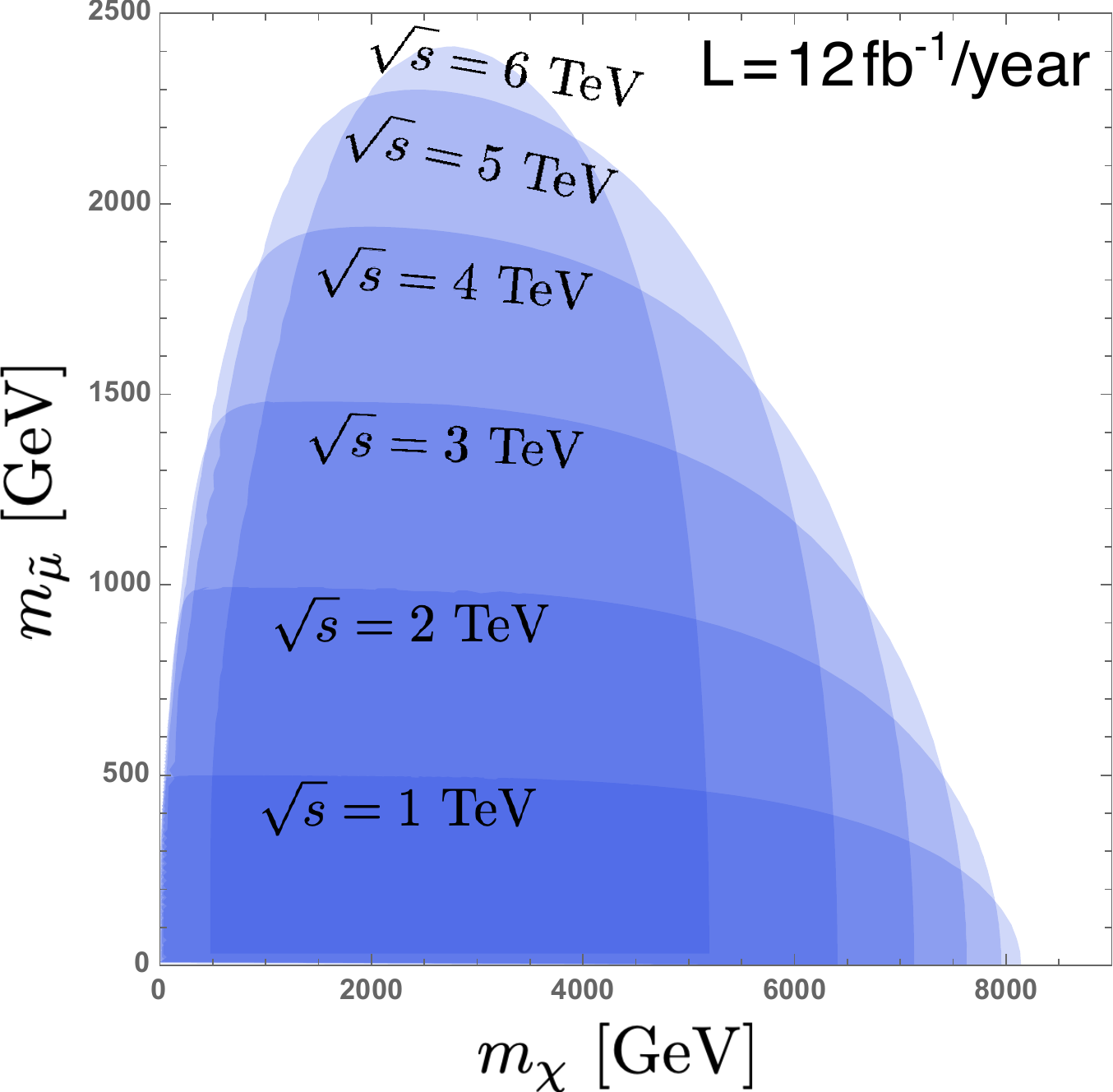}
\caption{The region where the number of the charged slepton pair
production per year is larger than a hundred at the $\mu^+ \mu^+$
collider. We take the luminosity to be ${\cal L} = 12~{\rm
fb}^{-1}/{\rm year}$. The beam polarizations are taken to be
$P_{\mu^+} = 0.8$ for both $\mu^+$ beams.}
\label{fig:susy_mumu}
\end{figure}
%%%%%%%%%%%%%%%%%%%%

At the $\mu^+ \mu^+$ collider, the kinematic reach of the scalar
lepton production is higher while the possible luminosity is reduced
in the current technology. At this collider, the neutralino exchange
provides a pair production of the scalar muons while scalar neutrino
productions are forbidden by charge conservation. 
The process of the scalar lepton production at the same sign muon
collider has been studied in Ref.~\cite{Heusch:1995yw}. A similar
process at $e^- e^-$ colliders has also been considered in
Refs.~\cite{Keung:1983nq,Cuypers:1993vy}. Non-vanishing amplitude
requires the intermediate neutralino in the $t$ and $u$-channel to be
a Majorana fermion, and thus the amplitude is proportional to the
gaugino mass. The differential cross sections is given by
\begin{align}
  d \sigma = {d \cos \theta \over 32 \pi} {\beta \over s} | M_{\rm RR} |^2
  {(1 + P_{\mu 1}) (1 + P_{\mu 2}) \over 4}, \quad 0 \leq \cos \theta \leq 1, 
\end{align}
where
\begin{align}
  M_{\rm RR} = - {g_2^2 \over 2} \cdot {4 \sqrt x_A (1 + 2 x_A - 2 x_3) 
   \over
  (1 + 2 x_A - 2 x_3)^2 - \beta^2 \cos^2 \theta
    },
\end{align}
and
\begin{align}
  x_A = {m_\chi^2 \over s}, \quad
  x_3 = {m_{\tilde \mu}^2 \over s}, \quad
  \beta = \sqrt { 1 - 4 x_3  }.
\end{align}
The production angle $\theta$ can be integrated to give the total
cross section as
\begin{align}
  \sigma = &{g_2^4 \over 64 \pi} {1 \over s} 
  \left[ {\beta x_A \over x_A + (x_A - x_3)^2}
  + {2 x_A \over 1 + 2 x_A - 2 x_3} 
  \log {{1+2x_A - 2 x_3 + \beta \over 1+2x_A - 2 x_3 - \beta}}
  \right]
  \nonumber \\
  & \times 
  {(1 + P_{\mu 1}) (1 + P_{\mu 2}) \over 4}.
\end{align}

We show in Fig.~\ref{fig:susy_mumu} the number of events per year by
assuming the luminosity, ${\cal L} = 12$~fb$^{-1}$~year$^{-1}$, which
corresponds to 70~\% of running with Eq.~\eqref{eq:mumu_lumi}. Even
with very heavy gauginos, such as 5~TeV, the yearly production rate
can easily be over a hundred.
Although the luminosity is lower compared to the $\mu^+ e^-$
colliders, the reach is much better.
This can be understood by the different $\sqrt s$ dependence of the
cross section.
For $m_\chi \gtrsim \sqrt s$, one
can integrate out the gaugino, and we obtain effective contact
interactions. The $(\bar \mu \gamma_\mu e) (\tilde e^* \partial^\mu
\tilde \mu)$ term for the $\tilde e^- \tilde \mu^+$ production has 
mass dimension of six, whereas the $\tilde \mu^+ \tilde \mu^+$ production
occurs through the $(\bar \mu \mu^c)(\tilde \mu \tilde \mu)$ term, whose mass dimension is five. This makes the $\tilde \mu^+ \tilde
\mu^+$ productions less suppressed for large gaugino masses. On the
other hand, since the amplitude vanishes when the gaugino mass goes to
zero, the small gaugino mass region is not fully covered.
One can see that for $\sqrt s = 6~{\rm TeV}$ the reach starts to
shrink for a fixed luminosity.

The searches should cover interesting parameter regions motivated by
the muon $g-2$ anomaly~\cite{Muong-2:2006rrc, Muong-2:2021ojo}. 
In order to explain the discrepancy between the Standard Model
predictions and the experiments by the contributions from the
superparticles, their masses need to be less than about TeV. For a
recent study, see Ref.~\cite{Endo:2021zal}.

\section{Summary}
\label{sec:summary}

The muon collider is not just one of the options in future collider
experiments. It would be the only option of energy frontier in the
future as the electron or proton accelerations will soon reach the
limit in terms of the size of the experiments. The muon colliders
should in principle be possible, and if it realizes, physics impact is
quite strong. In addition to direct reach to $O(10)$ TeV energies,
they simultaneously provide good precision physics such as the Higgs
boson couplings. Not only collider physics, rich muon and neutrino physics
programs will be possible at the facility of the muon collider. 
Fortunately, we do have a technology of handling $\mu^+$ already to
form accelerator beams. 
This means that $\mu^+ e^-$ and $\mu^+ \mu^+$ colliders
are realistic options at the present time and worth considering and planning seriously.

We estimated the luminosity of the $\mu^+ e^-$ and $\mu^+ \mu^+$
colliders with realistic accelerator parameters, and demonstrated that
the $\mu^+ e^-$ colliders serve as very good Higgs boson factories.
For the $\mu^+ \mu^+$ colliders, the luminosity is reduced in the
current technology. Nevertheless, it is already found to be good
enough to search for new particles such as superparticles.

There are numbers of works to be done. One should consider the design
of the detectors and study physics capabilities based on it. The
measurements of the Higgs boson couplings can be done by using decay
modes other than $h \to b \bar b$ such as $h \to WW^*$, $\tau^+
\tau^-$, and $ZZ^*$.
Various new physics models other than supersymmetry can be searched at
$\mu^+ e^-$ and $\mu^+ \mu^+$ colliders. The option of $\gamma \mu^+$
colliders may also be interesting to explore since that would directly
prove the microscopic physics to explain the muon $g-2$ anomaly.

The problem of the neutrino radiations from the muon decay in the
beam also needs to be seriously considered~\cite{King:1999ja}.
Although the problem is less severe than 10~TeV class muon
colliders, going into deep underground and/or a careful design of the
beam transportation such as a tilt in the accelerating and interaction
regions may be needed.

\section*{Acknowledgements}
We would like to thank Takayuki Yamazaki, Cedric Zhang, and Shusei Kamioka for providing us with information of the muon yields on which our estimates are based.
The work is supported by JSPS KAKENHI Grant Nos.~JP19H00689~(RK, RM),
JP19K14711~(HT), JP21H01086~(RK), JP21J01117~(YH) and MEXT KAKENHI
Grant No.~JP18H05542~(RK, HT).

\appendix
\numberwithin{equation}{section}
\setcounter{equation}{0}

\bibliography{bibcollection}
\bibliographystyle{hyperieeetr2}

\end{document}